\documentclass[journal]{new-aiaa}
\usepackage{graphicx}
\usepackage{xcolor}
\usepackage{amsmath}
\usepackage{amsthm}
\usepackage{subcaption}
\usepackage{tikz}
\usetikzlibrary{arrows.meta,calc}

\newtheorem{theorem}{Theorem}

\newtheorem{lemma}{Lemma}
\newtheorem{proposition}{Proposition}
\newtheorem{assumption}{Assumption}
\theoremstyle{definition}
\newtheorem{definition}{Definition}
\newtheorem{problem}{Problem}
\theoremstyle{remark}
\newtheorem{remark}{Remark}
\theoremstyle{plain}

\allowdisplaybreaks

\title{Finite-Time Capture by a Dubins Pursuer Against Unknown Evader Strategies}

\author{Dinesh Patra\footnote{Ph.D. student, Department of Electrical Engineering; dinesh.patra912@kgpian.iitkgp.ac.in (Corresponding Author).},
Bhargav Jha\footnote{Assistant Professor, Department of Electrical Engineering; bhargav@ee.iitkgp.ac.in (Corresponding Author).},
and Ashish R. Hota\footnote{Associate Professor, Department of Electrical Engineering; ahota@ee.iitkgp.ac.in.}}
\affil{Indian Institute of Technology Kharagpur, Kharagpur, West Bengal 721302, India}

\begin{document}

\maketitle

\begin{abstract}
We consider a planar pursuit-evasion problem in which a Dubins pursuer seeks to capture a slower first-order point-mass evader that can change its heading instantaneously. First, we propose a turn-rate feedback law with a single tunable gain and derive an explicit admissible gain interval that guarantees finite-time capture against any admissible evader strategy while satisfying the turn-rate constraint. Next, we obtain an explicit capture-time bound and characterize the turning authority required for the gain interval to be non-empty. Finally, we show that the dependence on the evader's speed bound can be removed by deriving a speed ceiling, computable solely from the pursuer's parameters, below which the guarantees hold at the maximum allowable gain. Numerical simulations illustrate the effectiveness of the proposed approach.
\end{abstract}

\section{Introduction} \label{section:introduction}

 \lettrine{F}{ixed-wing} aircraft, wheeled robots, and other vehicles that move at a nearly constant speed and steer through a bounded turn rate are routinely tasked with capturing a moving target, in applications ranging from aerial combat \cite{ref_aerial_xu2026adaptive,shima2009uav,shinar1980three,shinar1981solution} and surveillance \cite{ref_survillience_lozano2022surveillance,matveev2017tight} to target tracking \cite{ref_target_tracking_zou2018optimal} and formation control \cite{bruckstein1993ant,marshall2004formations,tripathy2024convergence,ref_formation_zhou2023distributed}; see \cite{shneydor1998missile} for an overview of the underlying guidance problems. When the target is an adversarial evader, a practical guidance law for such a pursuer must meet three requirements. First, it must guarantee capture against every evader strategy while using only quantities available to the pursuer. Second, its turn-rate command must never exceed the actuator limit. Third, the command should vary continuously in time rather than switch between extreme values. This paper presents a feedback law that meets all three.

Much of the existing theory treats simpler settings. Pursuit games between holonomic players with simple motion, who can change their direction of motion instantaneously, are well developed \cite{ref_isaacs1999differential}. 
Open-loop capture strategies are known for one-on-one \cite[Chapter~6]{ref_isaacs1999differential}, multiplayer \cite{pachter2019two}, and obstacle-cluttered \cite{oyler2016pursuit} engagements. 
The singular surfaces that arise in these solutions are examined in \cite{dorothy2024one}, which proposes a Lyapunov-based feedback strategy that guarantees capture within an $\varepsilon$-neighborhood of the initial Apollonius circle \cite[Chapter~6]{ref_isaacs1999differential}.

Cars, robots, and aircraft, however, cannot turn instantaneously. A constant-speed unicycle with zero minimum turn radius captures this nonholonomic constraint alone, and the Markov--Dubins model \cite{Dubins1957,LaValle2006,sussmann1995shortest} adds a bounded turn rate, which makes the minimum turn radius a central parameter. For both models, time-optimal open-loop interception has been characterized only when the target is stationary \cite{chen2019nonlinear} or follows a trajectory known in advance \cite{wang2026nonlinear,meyer2015dubins,zheng2021time}, and the resulting solutions are not available in closed form.

Closed-form guidance laws are straightforward to implement, but their guarantees rely on benign target motion. Linearized proportional navigation \cite{zarchan2012tactical,kreindler1973optimality}, impact-angle-constrained guidance \cite{shaferman2008linear,ratnoo2008impact}, Lyapunov-based steering in polar coordinates \cite{ref_aicardi1995closed}, and trajectory-shaping guidance for path following \cite{ref_ratnoo2015path} are all analyzed for stationary or non-maneuvering targets. For a constant-speed unicycle, the bearing-proportional turn-rate law of \cite{ref_ttimp_dovrat2021tracking} ensures only that the pursuer either captures the target or settles into a bounded tracking motion. None of these laws, therefore, meets the first requirement.

The homicidal chauffeur game (HCG) \cite{ref_isaacs1999differential,ref_hcg_merz1974homicidal,ref_hcg_pachter2019classical} addresses the adversarial setting directly, where a faster pursuer with bounded curvature seeks to capture a slower evader that can change its heading instantaneously. Its solution is intricate because the type and arrangement of singular surfaces, and hence the structure of the optimal strategies, change considerably across parameter regimes \cite{ref_hcg_merz1974homicidal}. The game has been extended to bounded controls for both players, both under kinematics linearized about the collision triangle \cite{shinar1981solution} and in three dimensions \cite{shinar1980three}; to cooperative and reach--avoid variants \cite{bopardikar2009cooperative,hcg_tac_23_yan2023homicidal,hcg_automatica_24_yan2024multiplayer}; and to persistent tracking \cite{ref_gt_tracking_zhang2014game}, where the objective is to keep the target within a sensing region rather than to capture it, and the target must start inside a computed viability set.

 The HCG pursuer, however, violates the third requirement and, because its strategy depends on the evader’s speed, partly the first. 
 Under the optimal HCG strategy, the pursuer turns at the maximum rate until its heading is aligned with the line of sight (LOS) and then switches abruptly to straight-line motion.
 The command is therefore saturated throughout the turn and discontinuous at the switch, where measurement noise may cause chattering, thereby accelerating actuator deterioration.
 Bounded-curvature-rate maneuvers that avoid such switching have been constructed \cite{bakolas2009generation,rao2024curvature,shanmugavel2010co}, but only for stationary targets or targets with known trajectories. In addition, HCG winning conditions and strategies are expressed in terms of the pursuer--evader speed ratio, so the evader's maximum speed must either be known or be estimated online. In an HCG where the evader may switch once from a lower to a higher speed, \cite{ref_mahapatra2025deception} shows that a pursuer estimating this speed from the evader's observed motion plays suboptimally and that from certain initial configurations the evader can prolong capture by initially moving slowly and later revealing its higher speed; moreover, a conservative estimate close to the pursuer's own speed does not benefit the pursuer.

 A different route was taken in \cite{ref_krstic2026directional,ref_krystic_arxiv_wang2026nonholonomic}, where finite-time directional capture is established using input-to-state stability-based feedback designs. These designs use the evader's heading and speed and allow the pursuer to vary its forward speed, while \cite{ref_krystic_arxiv_wang2026nonholonomic} additionally requires a known bound on the evader's turning rate.

Consequently, existing methods either assume known target maneuvers, require information that is unavailable to the pursuer, such as the evader's heading or maximum speed, or yield saturated, discontinuous turn-rate commands. 
Thus, to the best of our knowledge, the existing literature does not provide a closed-form feedback law for a Dubins pursuer that guarantees finite-time capture of a simple-motion evader with a continuous, turn-rate-feasible command without relying on the evader's strategy in its design.

 In view of the above limitations, this paper proposes a turn-rate feedback law with a single gain that uses only the range to the evader and the pursuer's heading error relative to the LOS. Its contributions are as follows.
\begin{enumerate}
  \item An explicit interval of gains is derived for which, against every admissible evader strategy, capture occurs in finite time and the turn-rate command remains continuous in time and within its limit. Furthermore, an explicit upper bound on the capture time is obtained.
  \item The turning authority the pursuer needs for this interval to be non-empty is characterized in terms of its speed, its capture radius, and the evader's speed bound.
  \item A speed ceiling $\mu^\star$, computable from the pursuer's parameters alone, is derived such that the largest admissible gain guarantees capture and turn-rate feasibility against every evader slower than $\mu^\star$. The gain can therefore be designed without knowing or estimating the evader's speed.
\end{enumerate}
Numerical simulations against representative evader strategies validate these guarantees.

The remainder of this paper is organized as follows. Section \ref{section:problem_description} presents the kinematics, the information available to the pursuer, and the problem statement. Section \ref{section:proposed_law} presents the proposed law and the main results, Section \ref{section:numerical} reports numerical results, and Section \ref{section:conclusion} concludes the paper.
%%%%%%%%%%%%%%%%%%%%%%%%%%%%%%%%%%%%%%%%%%%%%%%%%%%%%%%%%%%%%%%%%%%%
%%%%%%%%%%%%%%%%%%%%%%%%%%%%%%%%%%%%%%%%%%%%%%%%%%%%%%%%%%%%%%%%%%%%
%%%%%%%%%%%%%%%%%%%%%%%%%%%%%%%%%%%%%%%%%%%%%%%%%%%%%%%%%%%%%%%%%%%%

\section{Problem Formulation}\label{section:problem_description}

We consider a two-agent pursuit-evasion problem in the unbounded planar domain $\mathbb{R}^2$. The pursuer is a non-holonomic vehicle following Dubins kinematics, while the evader is modeled as a point mass with first-order kinematics. The instantaneous positions of the pursuer and the evader are denoted by $p = (x_p, y_p)$ and $e= (x_e, y_e)$, respectively. The pursuer moves with a constant forward speed, denoted by $v > 0$, and its control input is given by its turning rate, denoted by $\omega_p$, which needs to satisfy $|\omega_p|\le\omega_m$. The turning rate controls its heading, which is denoted by $\theta$. Meanwhile, the evader's control inputs constitute its speed, denoted by $\mu \in [0,\mu_m]$, and its heading angle, denoted by $\psi$. The states of the pursuer and the evader are denoted by ${z_p}:=[x_p, y_p, \theta]^\top$ and ${z_e}:=[x_e, y_e]^\top$, respectively, and these states evolve as
\begin{subequations} \label{eq:kinematics}
\begin{align}
    &\ \dot{x}_p = v \cos \theta, \; \; %v_i \cos \psi_i
\dot{y}_p = v \sin \theta, \; \; \dot{\theta}= \omega_p, \label{eq:P_dyn}
\\ &\		\dot{x}_e =\mu \cos \psi, \;
		\dot{y}_e =\mu \sin \psi,  \label{eq:E_dyn}
\end{align}    
\end{subequations}
where the dependence on time $t$ is suppressed for better readability. The angles $\theta,\psi\in\mathbb{S}^{1}$. Throughout the paper, we assume that $\mu_m<v$. 

\begin{figure}[t!]
\centering
\begin{tikzpicture}[>={Stealth[length=2.2mm]}, font=\small]
  % geometry: heading theta = 27 deg, LOS angle beta = atan2(3.0, 2.4) = 51.34 deg
  \coordinate (P) at (2.2,2.3);   % pursuer
  \coordinate (E) at (4.6,5.3);   % evader
  % axes
  \draw[->, thick] (0,0) -- (6.2,0) node[right] {$x$};
  \draw[->, thick] (0,0) -- (0,6.2) node[above] {$y$};
  % projection lines
  \draw[dashed, thick] (0,5.3) node[left] {$y_e$} -- (E);
  \draw[dashed, thick] (4.6,0) node[below] {$x_e$} -- (E);
  \draw[dashed, thick] (0,2.3) node[left] {$y_p$} -- (4.9,2.3);
  \draw[dashed, thick] (2.2,0) node[below] {$x_p$} -- (P);
  % angle arcs centred at the pursuer
  \draw[blue!70!black, thick] ($(P)+(0:1.2)$) arc[start angle=0, end angle=27, radius=1.2];
  \node[blue!70!black] at ($(P)+(13.5:1.45)$) {$\theta$};
  \draw[orange!50!black, line width=1.2pt] ($(P)+(27:1.6)$) arc[start angle=27, end angle=51.34, radius=1.6];
  \node[orange!50!black] at ($(P)+(39.2:1.85)$) {$\alpha$};
  \draw[black!85, line width=1.2pt] ($(P)+(0:2.2)$) arc[start angle=0, end angle=51.34, radius=2.2];
  \node[black!85] at ($(P)+(32:2.5)$) {$\beta$};
  % line of sight and range
  \draw[red!60!black, line width=1.3pt, dash pattern=on 1.5pt off 1.8pt] (P) -- (E)
    node[pos=0.62, above, sloped, text=red!60!black] {$\rho(t)$};
  % pursuer body rotated by theta
  \begin{scope}[shift={(P)}, rotate=27]
    \draw[blue!70!black, very thick, rounded corners=5pt] (-0.6,-0.55) rectangle (0.6,0.55);
    \draw[fill=gray!25, thick, rounded corners=2pt] (-0.45,0.6) rectangle (0.45,0.78);
    \draw[fill=gray!25, thick, rounded corners=2pt] (-0.45,-0.78) rectangle (0.45,-0.6);
  \end{scope}
  % pursuer heading (velocity)
  \draw[->, blue!70!black, thick] (P) -- ++(27:3.0) node[right] {$v$};
  \fill[blue!70!black] (P) circle (1.5pt) node[below left=1pt] {$p$};
  % evader and its velocity
  \fill[red!85!black] (E) circle (2.5pt) node[above right=0pt] {$e$};
  \draw[->, red!85!black, thick] (E) -- ++(150:0.9) node[above left=-1pt] {$\mu$};
\end{tikzpicture}
\caption{Relative geometry between a Dubins pursuer and a point-mass evader, illustrating the range ($\rho$) separating them, and the angles $\beta$ and $\alpha$.}
\label{fig:unicycle_pursuit}
\end{figure}
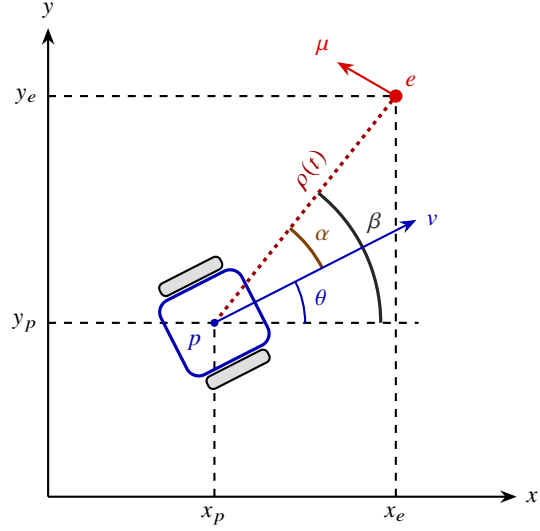

Based on the relative geometry between the pursuer and the evader, we define
\begin{subequations} \label{eq:relative_kinematics}
\begin{align}
\rho(t) & := \|e(t)-p(t)\|_2, \label{eq:dist}\\
\beta(t) & := \operatorname{atan2}\!\big(y_e(t)-y_p(t),\; x_e(t)-x_p(t)\big), \label{eq:beta}\\
\alpha(t) & := \operatorname{atan2}\!\big(\sin(\beta(t)-\theta(t)),\;
              \cos(\beta(t)-\theta(t))\big), \label{eq:alpha}
\end{align}    
\end{subequations}
where $\rho(t)$ is the distance or range separating the two agents, $\beta(t)$ is the
line-of-sight (LOS) angle from the pursuer to the evader, and $\alpha(t)$ is the heading error, i.e., the signed angular deviation of the pursuer's heading from the LOS direction, measured along the shorter arc and considered positive counterclockwise. Consequently, $\alpha(t)$ satisfies $|\alpha(t)|\le\pi$, i.e., $\alpha(t) \in (-\pi,\pi]$. These quantities are illustrated in Fig. \ref{fig:unicycle_pursuit}.

%%%%%%%%%%%%%%%%%%%%%%%%%%%%%%%%%%%%%%%%%%%%

We now present some key definitions and structural assumptions required for our analysis. 

\begin{definition}[Finite-Time Capture] \label{def:capture}
The evader is said to be \emph{captured in finite time} if the first instant at which the range equals the capture radius $r_c > 0$, and is strictly decreasing, is finite. In other words, the time of capture $$t_c:=\inf\{\,t\ge0: \rho(t)=r_c, \ \dot\rho(t)<0\,\}<\infty.$$
\end{definition}

The above notion of capture is analogous to the setting of the classical HCG \cite{ref_hcg_pachter2019classical}. We note that the engagement is defined on $[0,t_c]$ and terminates at $t_c$, which may correspond to the physical neutralization of the evader; thus the kinematics \eqref{eq:kinematics} are not considered beyond that instant. The engagement is considered only up to $t_c$, and thus, over the interval $t \in [0,t_c]$, the range $\rho(t)$ satisfies $\rho(t)\ge r_c$, with $\rho(0)>r_c$. We now impose the following assumptions. 

\begin{assumption}[Information Structure]\label{ass:info}
At each time $t \geq 0$, the pursuer has access to the range $\rho(t)$, the heading error $\alpha(t)$, and the upper bound\footnote{Subsequently, in Proposition \ref{prop:mustar}, we show that our approach extends to the case when $\mu_m$ is not known and the guarantees on capture and turn-constraint satisfaction hold against every evader whose speed bound $\mu_m$ lies below a ceiling, which can be computed a priori using only the pursuer's parameters.} $\mu_m$ on the evader's speed. Its own parameters $v$, $\omega_m$, and $r_c$ are known. However, the evader's instantaneous speed $\mu$ and its heading angle $\psi$ remain unknown to the pursuer.
\end{assumption}

\begin{assumption}[Admissible Evader Strategy] \label{ass:evader_strategy}
An evader strategy is said to be admissible if $(\mu(t),\psi(t))$ are such
that \eqref{eq:E_dyn} admits a solution in the Carath\'eodory sense \cite{ref_cortes2008discontinuous},
and satisfy $\mu(t)\le\mu_m<v$.
\end{assumption}

Assumption \ref{ass:evader_strategy} permits discontinuous strategies with the derivatives in \eqref{eq:E_dyn} holding for almost every $t$. No further assumptions are imposed on the evader's strategy. In particular, the evader may employ any random, time-varying, feedback, or open-loop policy, including any worst-case strategy aimed at delaying capture. We now state the problem that we investigate in this paper.

\begin{problem}\label{prob:main}
Given any initial configuration of the pursuer and evader with $\rho(0)>r_c$,
our goal is to design a control strategy for the pursuer's turn rate $\omega_p$ that respects the actuator limit $|\omega_p|\le\omega_m$ and guarantees finite-time capture
of the evader under Assumptions \ref{ass:info} and \ref{ass:evader_strategy},
regardless of the admissible strategy employed by the evader. 
\end{problem}

%%%%%%%%%%%%%%%%%%%%%%%%%%%%%%%%%%%%%%%%%%%%%%%%%%%%%%%%%%%%%%%%%%%%
%%%%%%%%%%%%%%%%%%%%%%%%%%%%%%%%%%%%%%%%%%%%%%%%%%%%%%%%%%%%%%%%%%%%

\section{Control Strategy for Finite-Time Capture}\label{section:proposed_law}

%%%%%%%%%%%%%%%%%%%%%%%%%%%%%%%%%%%%%%%%%%%%%%%%%%%%%%%%%%%%%%%%%%%%
%%%%%%%%%%%%%%%%%%%%%%%%%%%%%%%%%%%%%%%%%%%%%%%%%%%%%%%%%%%%%%%%%%%%

In this section, we present the proposed control strategy for the pursuer and prove that it achieves finite-time capture. First, we compute the time derivatives of the range and the heading error defined in \eqref{eq:dist} and \eqref{eq:alpha}, respectively, by leveraging the kinematics from \eqref{eq:kinematics}, as
\begin{subequations} \label{eq:der_relative_kinematics}
\begin{align}
\dot{\rho} &= \mu\cos(\psi-\beta) - v\cos\alpha, \label{eq:rho_dot}\\
% \dot{\beta} &= \frac{\mu\sin(\psi-\beta) + v\sin\alpha}{\rho}, \label{eq:beta_dot}\\
\dot{\alpha} &= \frac{\mu\sin(\psi-\beta) + v\sin\alpha}{\rho} - \omega_p. \label{eq:alpha_dot}
\end{align}    
\end{subequations}

Since $\rho(t)\ge r_c>0$ on $[0,t_c]$, the range is continuous and differentiable. The expression \eqref{eq:alpha_dot} is valid whenever $\alpha$ is differentiable, i.e., except when\footnote{Theorem \ref{thm:up_attract}(a) shows that no such instance arises under the subsequently proposed control law with a sufficiently large gain.} $\alpha \in (-\pi,\pi]$ jumps across $+\pi$ and $-\pi$ by $2\pi$. We note from \eqref{eq:rho_dot} that the expression for $\dot\rho$ is independent of $\omega_p$ and depends only on the heading error $\alpha$. The following lemma makes this precise by characterizing the range of $\alpha$ for which the evader can prevent the range from decreasing.

%%%%%%%%%%%%%%%%%%%%%%%%%%
\begin{lemma}[Range-decrease condition]\label{lem:range-hold}
Let $\gamma=\mu_m/v\in[0,1)$ denote the speed ratio. At any instant $t$, and irrespective of the turn rate applied by the pursuer, the evader can enforce $\dot\rho(t)\ge0$ through an admissible choice of its inputs  if and only if
\begin{equation}\label{eq:range_hold}
  |\alpha(t)|\;\ge\;\cos^{-1}\gamma .
\end{equation}
In particular, if $|\alpha(t)|<\cos^{-1}\gamma$, then $\dot\rho(t)<0$ for every
admissible evader strategy.
\end{lemma}
%%%%%%%%%%%%%%%%%%%%

\begin{proof}
We suppress the argument $t$ in the proof for better readability. Maximizing $\dot\rho$ from \eqref{eq:rho_dot} with respect to the admissible inputs $\mu\in[0,\mu_m]$ and $\psi$ yields $\max_{\{\mu\le\mu_m,\,\psi\}}\dot\rho=\mu_m-v\cos\alpha$, attained at $\mu=\mu_m$ and $\psi=\beta$. 

Hence, the evader can enforce $\dot\rho\ge0$ if and only if $\mu_m-v\cos\alpha\ge0$, i.e., $\cos\alpha\le\gamma$. As cosine is an even function, this results in $\cos\alpha=\cos|\alpha|\le\gamma$, and since $|\alpha|\le\pi$ following the definition of $\alpha$ in \eqref{eq:alpha} and cosine
is strictly decreasing on $[0,\pi]$, we obtain $|\alpha|\ge\cos^{-1}\gamma$. Conversely, $|\alpha|<\cos^{-1}\gamma$ implies $\cos\alpha>\gamma$ and therefore $\dot\rho\le\mu_m-v\cos\alpha<0$ for every admissible evader strategy. This completes the proof.
\end{proof}

%%%%%%%%%%%%%%%%%%%%%%%%%%
Accordingly, we denote the set of heading errors in the interior of which the range is guaranteed to decrease by
\begin{equation}\label{eq:Pset}
\mathcal{P} := \{\, \alpha \in (-\pi,\pi] \mid -\cos^{-1}\gamma \le \alpha \le \cos^{-1}\gamma \,\}.
\end{equation}
Since the inverse cosine has principal range $[0,\pi]$ and $\gamma\in[0,1)$,
we have $\cos^{-1}\gamma\in(0,\pi/2]$, so $\mathcal{P}\subseteq[-\pi/2,\pi/2]$
has nonempty interior. Having characterized the values of the heading error at which the range is
guaranteed to decrease, we now propose the turn-rate control law
\begin{equation}\label{eq:thm_controller}
  \omega_p := \frac{v\sin\alpha}{\rho} + k\alpha,
\end{equation}
where the gain $k>0$ is a design parameter. Following Assumption \ref{ass:info}, the control law \eqref{eq:thm_controller} depends on the relative geometry only through $\rho$ and $\alpha$, which are typically available directly from a body-fixed sensor. In particular, it does not require the evader's instantaneous speed or heading angle. 

We note that the pursuer's turn rate is limited to $\omega_m$. Since $|\alpha|\le\pi$ by the definition of $\alpha$ in
\eqref{eq:alpha} and $\rho\ge r_c$ on $[0,t_c]$, the commanded turn rate in
\eqref{eq:thm_controller} is bounded, and the actuator limit translates directly into an
upper bound on the admissible gain $k$, as established next.

\begin{lemma}[Satisfaction of the turn constraint]\label{lem:turn}
Let the pursuer apply the control law \eqref{eq:thm_controller} with a gain $k\in(0,k_{\max}]$, where
\begin{equation}\label{eq:k_max}
  k_{\max} \;:=\; \frac{\omega_m-\frac{v}{r_c}}{\pi} .
\end{equation}
Then the turn constraint $|\omega_p|\le\omega_m$ is satisfied for every admissible evader strategy and every configuration with $\rho\ge r_c$.
\end{lemma}
%%%%%%%%%%%%%%%%%%%%%%%%%%%%%%%%%%%%%%%%%%%%%
\begin{proof}
From \eqref{eq:thm_controller} and the triangle inequality,
\begin{equation}\label{eq:bound_omega}
  |\omega_p| = \left|\frac{v\sin\alpha}{\rho}+k\alpha\right|
  \le \frac{v|\sin\alpha|}{\rho}+k|\alpha|
  \le \frac{v}{r_c}+k\pi ,
\end{equation}
where the last inequality follows from $|\sin\alpha|\le1$, $\rho\ge r_c$, and
$|\alpha|\le\pi$. Hence $|\omega_p|\le\omega_m$ whenever
$v/r_c+k\pi\le\omega_m$, which rearranges to \eqref{eq:k_max}.
\end{proof}

Lemma \ref{lem:turn} bounds the admissible gain from above, as a direct consequence of the actuator limit alone. A lower bound on $k$, in contrast, arises from the requirement to drive the heading error into the set $\mathcal{P}$ defined in \eqref{eq:Pset}. Substituting the proposed control law \eqref{eq:thm_controller} into \eqref{eq:alpha_dot} yields
\begin{equation}\label{eq:alpha_cl}
\dot{\alpha} = -k\alpha + d(t), \, \text{where} \, \, d(t) = \frac{\mu\sin(\psi-\beta)}{\rho}.
\end{equation}
Since $\mu \leq \mu_m$, $|\sin(\psi-\beta)| \leq 1$, and $\rho(t) \geq r_c$ for
all $t \in [0,t_c]$, the evader term is bounded by
\begin{equation}\label{eq:d_bound}
|d(t)| \leq \frac{\mu_m}{r_c}.
\end{equation}

%%%%%%%%%%%%%%%%%%%%%%
%%%%%%%%%%%%%%%%%%%%%%%%%%%%%%%%%%%%%%%%
We now analyze the closed-loop dynamics \eqref{eq:alpha_cl} and the evolution of the heading error in the following theorem. To this end, we define
\begin{equation}\label{eq:kmin_def}
k_{\min} \;:=\; \frac{\mu_m}{r_c}\sqrt{\frac{v}{2(v-\mu_m)}}, \quad A^\star:=\mu_m/(kr_c),
\end{equation}
where $v>\mu_m$ ensures that $k_{\min}$ is well defined, with $k_{\min}>0$ and $A^\star>0$ for $\mu_m>0$.
%%%%%%%%%%%%%%%%%%%%%%%%%%%%

\begin{theorem}[Convergence of the heading error and finite-time capture]
\label{thm:up_attract}
Consider the kinematics \eqref{eq:kinematics} under Assumptions \ref{ass:info} and \ref{ass:evader_strategy} and let $k_{\min} < k_{\max}$. Suppose the pursuer applies the control law \eqref{eq:thm_controller} with a gain $k \in (k_{\min} , k_{\max} ]$. 
Then, for every admissible evader strategy and every initial configuration with $\rho(0)>r_c$ and $\alpha(0)\in(-\pi,\pi]$, the following are true. 
\begin{enumerate}
\item[(a)] $\mathcal{M} := \{\, \alpha \in (-\pi,\pi] \mid -A^\star \le \alpha \le A^\star \,\}$ is the minimal positively invariant set containing the origin with respect to $\alpha$ and is globally exponentially attractive, i.e., for every $t \in [0,t_c]$,
\begin{equation}\label{eq:dist_bound}
  \operatorname{dist}\!\big(\alpha(t),\mathcal{M}\big) \;\le\;
  \operatorname{dist}\!\big(\alpha(0),\mathcal{M}\big)\, e^{-kt},
\end{equation}
where $\operatorname{dist}(\alpha(t),\mathcal{M}) = \max\{|\alpha(t)| - A^\star,\, 0\}$.
Furthermore, $\mathcal{M}$ is a proper subset of $(-\pi,\pi]$, and consequently, $\alpha$ does not jump across $\pm\pi$ and is continuous throughout the engagement, resulting in the continuity of the proposed turn-rate law \eqref{eq:thm_controller}.
\item[(b)] The heading error enters the set $\mathcal{P}$ in finite time, denoted by $\tau_p$, which satisfies 
\begin{equation}\label{eq:tauP}
\tau_{p} \;\le\; \bar{\tau}_p\;:=\;\frac{|\alpha(0)|-\cos^{-1}\gamma}{k\bigl(\cos^{-1}\gamma-A^\star\bigr)}, \qquad \text{if } |\alpha(0)|>\cos^{-1}\gamma. 
%\\[1ex]   
\end{equation}
Note that $\tau_{p} = 0$ if $|\alpha(0)|\le\cos^{-1}\gamma$. Once the heading error enters $\mathcal{P}$, it remains in its interior thereafter; in particular, $\mathcal{P}$ is positively invariant and contains $\mathcal{M}$ in its interior.
\item[(c)] Capture in the sense of Definition \ref{def:capture} occurs in finite time, and the capture instant satisfies
\begin{align}
    t_c & \;\le\; \bar{t}_c \;:=\; \frac{\rho(0)-r_c+C}{\delta(k)} , \label{eq:tcbound} \\
\text{where}  \, \, \, \delta(k)&:=(v-\mu_m)-\frac{v\mu_m^{2}}{2k^{2}r_c^{2}}, \label{eq:delta}\\
\text{and}  \, \, \,  C&:=\frac{v}{k}\left(\frac{(A^\star)\,|\alpha(0)|}{2}
                       +\frac{\alpha^{2}(0)}{4}\right) \geq 0, \label{eq:Cconst}
\end{align}
with $\delta(k)>0$ when $ k > k_{\min}$. Furthermore, $\bar\tau_{p}<\bar{t}_c$.
\end{enumerate}
\end{theorem}
%%%%%%%%%%%%%%%%%%%%%%%%%%%%%%%%%%%%%%%%%%%%%%%%%%%%%%%%%%%%%%%%%%%%%%%%%%%%
\begin{proof}
\emph{(a)} We first inspect the closed-loop evolution of the heading error from \eqref{eq:alpha_cl}, i.e., $\dot\alpha=-k\alpha+d(t)$, where the evader's influence is confined to the term  $d(t)=\mu\sin(\psi-\beta)/\rho$ that satisfies $|d(t)|\le \mu_m/r_c$ by
\eqref{eq:d_bound}. For $\alpha\neq0$, $|\alpha|$ is differentiable with
$\frac{d}{dt}|\alpha|=\mathrm{sgn}(\alpha)\,\dot\alpha$; substituting
$\dot\alpha$ from \eqref{eq:alpha_cl} and using
$\mathrm{sgn}(\alpha)\,\alpha=|\alpha|$, we obtain
\begin{equation}\label{eq:abs_alpha_dot}
  \frac{d}{dt}|\alpha|
  = -k|\alpha| + \mathrm{sgn}(\alpha)\,d(t)
  \;\le\; -k\big(|\alpha|-A^\star\big),
\end{equation}
where the inequality follows from $\mathrm{sgn}(\alpha)\,d(t)\le|d(t)|\le \frac{\mu_m}{r_c} = k A^\star$. Since only the bound $|d|\le \frac{\mu_m}{r_c}$ has been used, \eqref{eq:abs_alpha_dot} holds irrespective of the evader's strategy. Because $A^\star>0$ for $\mu_m>0$\footnote{If $\mu_m=0$, then $d\equiv0$, $\mathcal{M}=\{0\}$, and the subsequent claims hold trivially.}, the point $\alpha=0$ lies in the interior of $\mathcal{M}$. Consequently, \eqref{eq:abs_alpha_dot} is invoked below only at instants where $|\alpha|\ge A^\star>0$, at which it is valid. Thus, excluding $\alpha=0$ is justified.

The right-hand side of \eqref{eq:abs_alpha_dot} is strictly negative when  $|\alpha|>A^\star$ and vanishes at $|\alpha|=A^\star$ since $k>k_{\min}>0$.
At the boundary $|\alpha|=A^\star$ we therefore have $\frac{d}{dt}|\alpha|\le0$, so the vector field does not point outward and $\mathcal{M} = \{\, \alpha \in (-\pi,\pi] \mid -A^\star \le \alpha \le A^\star \,\}$ is positively invariant, following Nagumo's theorem \cite{ref_book_blanchini2008set,ref_book_cohen2023adaptive}.
Hence, $|\alpha|$ decreases strictly outside $\mathcal{M}$. 
To quantify this decrease, note that $A^\star$ is constant, so \eqref{eq:abs_alpha_dot} can be written as
$\frac{d}{dt}\big(|\alpha|-A^\star\big)\le-k\big(|\alpha|-A^\star\big)$. Integrating over $[0,t]$ yields
\begin{equation}\label{eq:exp_decay}
  |\alpha(t)|-A^\star \;\le\; \big(|\alpha(0)|-A^\star\big)e^{-kt}.
\end{equation}
If $|\alpha(0)|> A^\star$, then the right-hand side of \eqref{eq:exp_decay} is strictly positive, and thus it bounds $\max\{|\alpha(t)|-A^\star,0\}$, which is precisely
$\operatorname{dist}(\alpha(t),\mathcal{M})$; this gives \eqref{eq:dist_bound}. If $|\alpha(0)|\le A^\star$, positive invariance keeps $\alpha(t)\in\mathcal{M}$ implying $|\alpha(t)|\le A^\star$ and \eqref{eq:dist_bound} therefore holds trivially with both sides equal to zero. In either case, $\mathcal{M}$ is globally exponentially attractive, with trajectories starting outside approaching it at least as fast as $e^{-kt}$.

To establish minimality, we show that no interval $[a,b]$ with $a<0<b$ that
fails to contain $\mathcal{M}$ is positively invariant under every admissible
evader strategy. Suppose first that $b<A^\star$. Since $A^\star=\mu_m/(kr_c)$, this gives $\mu_m/(kb)>r_c$, so the set $(r_c,\mu_m/(kb))$ is non-empty. Consider a configuration with $\alpha=b>0$ and any range $\rho \in (r_c,\mu_m/(kb))$, and let the evader choose $\mu=\mu_m$ and $\psi=\beta+\pi/2$, which gives $d=\mu_m/\rho$. Then following \eqref{eq:abs_alpha_dot}, $\dot\alpha\big|_{\alpha=b}=-kb+\frac{\mu_m}{\rho}>0$, since $\rho<\mu_m/(kb)$. Thus, the heading error increases, and the trajectory leaves $[a,b]$ through its upper endpoint. Suppose instead that $a>-A^\star$. By the same reasoning $\mu_m/(k|a|)>r_c$; taking $\alpha=a<0$, any $\rho\in(r_c,\mu_m/(k|a|))$, and the admissible choice
$\mu=\mu_m$, $\psi=\beta-\pi/2$, which gives $d=-\mu_m/\rho$, following \eqref{eq:abs_alpha_dot} we obtain
$\dot\alpha\big|_{\alpha=a}=k|a|-\frac{\mu_m}{\rho}<0$ and thus the trajectory leaves $[a,b]$ through its lower endpoint. Hence every positively invariant interval with respect to $\alpha$ satisfies $a\le-A^\star$ and $b\ge A^\star$, i.e., contains $\mathcal{M}$, and $\mathcal{M}$ is therefore the minimal such set containing the origin.

We now observe that 
$$k > k_{\min} = \frac{\mu_m}{r_c}\sqrt{\frac{v}{2(v-\mu_m)}}
> \frac{1}{\sqrt2}\,\frac{\mu_m}{r_c} > \frac{\mu_m}{\pi r_c},$$
where the inequalities follow from $v/(v-\mu_m)>1$ and $\pi>\sqrt2$. As a result, $A^\star=\mu_m/kr_c<\pi$, and hence $\mathcal{M}$ is a proper subset of $(-\pi,\pi]$. Consequently, $|\alpha|=\pi$ lies strictly outside $\mathcal{M}$, and $\frac{d}{dt}|\alpha|<0$ at $|\alpha|=\pi$ since $\mathcal{M}$ is globally exponentially attractive following the previous discussion based on \eqref{eq:abs_alpha_dot}. Hence the vector field points strictly inwards at $|\alpha|=\pi$. Therefore, the heading error does not cross $\pm\pi$, and is continuous throughout the engagement. Consequently, the commanded turn rate \eqref{eq:thm_controller} is continuous in time, and the proposed control
law involves no switching or discontinuity.

\medskip
\noindent\emph{(b)} 
We first establish $\mathcal{M}\subset\mathcal{P}$, i.e., $A^\star<\cos^{-1}\gamma$, where $\gamma=\mu_m/v \in [0,1)$. Since $k>k_{\min}$ from \eqref{eq:kmin_def}, we have
\begin{equation}\label{eq:Astar_bound}
  A^\star=\frac{\mu_m}{kr_c}<\frac{\mu_m}{k_{\min}r_c}
  =\sqrt{\frac{2(v-\mu_m)}{v}}=\sqrt{2(1-\gamma)} .
\end{equation}
For $\gamma \in [0,1]$, let $g(\gamma):=\cos^{-1}\gamma-\sqrt{2(1-\gamma)}$, where inverse cosine has principal range $[0,\pi]$. Then, $g(1)=0$. For $\gamma\in[0,1)$, we have
\[
  g'(\gamma)=\frac{1}{\sqrt{1-\gamma}}
  \Big(\frac{1}{\sqrt2}-\frac{1}{\sqrt{1+\gamma}}\Big)<0 ,
\]
because $\sqrt{1+\gamma}<\sqrt2$ for $\gamma \in [0,1)$. Hence, $g$ is strictly decreasing on $[0,1)$ with $g(1)=0$, so that $g(\gamma)>0$ in this interval. This results in $\cos^{-1}\gamma>\sqrt{2(1-\gamma)}>A^\star$ where the last inequality follows from \eqref{eq:Astar_bound}. Furthermore, recalling that the inverse cosine has principal range $[0,\pi]$,
$\gamma\in[0,1)$ gives $\cos^{-1}\gamma\in(0,\pi/2]$, and thus $\mathcal{M}\subset\mathcal{P}\subset(-\pi,\pi]$.

Now, following the above discussion, whenever $|\alpha|\ge\cos^{-1}\gamma>A^\star$, the right-hand side of \eqref{eq:abs_alpha_dot} is decreasing in $|\alpha|$. Thus, we get
\begin{equation}\label{eq:uniform_decay}
  \frac{d}{dt}|\alpha| \;\le\; -k\big(|\alpha|-A^\star\big)
  \;\le\; -k\big(\cos^{-1}\gamma-A^\star\big)\;<\;0 .
\end{equation}
In particular, at $|\alpha|=\cos^{-1}\gamma$ the vector field points strictly inward, thus $\mathcal{P}$ is positively invariant with respect to $\alpha$ following Nagumo's theorem \cite{ref_book_blanchini2008set,ref_book_cohen2023adaptive}. 
Furthermore, starting from any $|\alpha(0)|>\cos^{-1}\gamma$, the rate of decrease of $|\alpha|$ is bounded below by a nonzero fixed rate, regardless of the evader's strategy. This implies $\alpha$ must enter $\mathcal{P}$ in finite time, denoted by $\tau_p$, and by integrating \eqref{eq:uniform_decay} from $t=0$ to $\tau_p$, we obtain
\begin{equation*}
\tau_p \le \dfrac{|\alpha(0)|-|\alpha(\tau_p)|}{k\bigl(\cos^{-1}\gamma-A^\star\bigr)} = \dfrac{|\alpha(0)|-\cos^{-1}\gamma}{k\bigl(\cos^{-1}\gamma-A^\star\bigr)}=:\bar{\tau}_p,    
\end{equation*}
where the equality follows from the fact that $\tau_p$ is defined as the time of entry of $\alpha$ into $\mathcal{P}$, starting from $|\alpha(0)|>\cos^{-1}\gamma$, i.e., $|\alpha(\tau_p)|=\cos^{-1} \gamma$. The case $|\alpha(0)|\le\cos^{-1}\gamma$ is trivial; the heading error starts in
$\mathcal{P}$ and $\tau_{p}=0<\bar{\tau}_p$. Figure \ref{fig:thm_1} illustrates the nested sets $\mathcal{M}\subset\mathcal{P}\subset(-\pi,\pi]$ and the evolution of the heading error from an initial value outside $\mathcal{P}$.

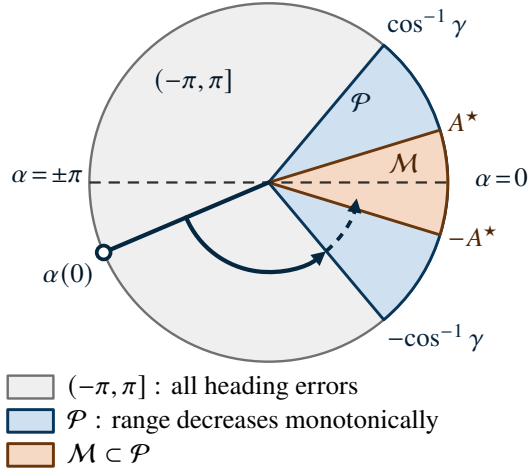
\begin{figure}[t!]
\centering
\begin{tikzpicture}[x=1pt, y=1pt, line join=round, font=\normalsize,
    colP/.style={draw={rgb,1:red,0.0314;green,0.2078;blue,0.3529}},
    colM/.style={draw={rgb,1:red,0.4157;green,0.2118;blue,0.0431}},
    colN/.style={draw={rgb,1:red,0.0196;green,0.1569;blue,0.2510}},
    txP/.style={text={rgb,1:red,0.0118;green,0.1255;blue,0.2275}},
    txM/.style={text={rgb,1:red,0.3020;green,0.1412;blue,0.0157}},
    txN/.style={text={rgb,1:red,0.0039;green,0.0784;blue,0.1373}}]
  % Geometry (radius R = 67.45 pt): P = [-50, 50] deg, M = [-17, 17] deg, alpha(0) at 203 deg
  % full set of heading errors (-pi, pi]
  \draw[line width=0.9pt, draw={rgb,1:red,0.4392;green,0.4392;blue,0.4392},
        fill={rgb,1:red,0.9372;green,0.9372;blue,0.9372}] (0,0) circle (67.45);
  % set P
  \draw[colP, line width=1.1pt, fill={rgb,1:red,0.8118;green,0.8823;blue,0.9412}]
    (0,0) -- (-50:67.45) arc[start angle=-50, end angle=50, radius=67.45] -- cycle;
  % set M
  \draw[colM, line width=1.1pt, fill={rgb,1:red,0.9647;green,0.8510;blue,0.7686}]
    (0,0) -- (-17:67.45) arc[start angle=-17, end angle=17, radius=67.45] -- cycle;
  % alpha = +-pi / alpha = 0 axis
  \draw[line width=0.85pt, draw={rgb,1:red,0.1804;green,0.1804;blue,0.1804},
        dash pattern=on 4.25pt off 2.975pt] (-67.45,0) -- (67.45,0);
  % initial heading error alpha(0)
  \draw[colN, line width=1.6pt, line cap=round] (0,0) -- (203:67.45);
  \draw[colN, line width=1.4pt, fill=white] (203:67.45) circle (2.6);
  % evolution of the heading error
  \draw[colN, line width=1.6pt, line cap=round,
        -{Triangle[length=5.5pt, width=5pt, fill={rgb,1:red,0.0196;green,0.1569;blue,0.2510}]}]
    (203:33.72) arc[start angle=203, end angle=310, radius=33.72];
  \draw[colN, line width=1.4pt, dash pattern=on 3.08pt off 2.8pt,
        -{Triangle[length=5.5pt, width=5pt, fill={rgb,1:red,0.0196;green,0.1569;blue,0.2510}]}]
    (310:33.72) arc[start angle=310, end angle=350, radius=33.72];
  % labels
  \node[txP, anchor=base west, inner sep=0pt] at (44.6, 56.7)  {$\cos^{-1}\gamma$};
  \node[txP, anchor=base west, inner sep=0pt] at (44.6, -61.7) {$-\!\cos^{-1}\gamma$};
  \node[txM, anchor=base west, inner sep=0pt] at (67.0, 19.5)  {$A^{\star}$};
  \node[txM, anchor=base west, inner sep=0pt] at (67.4, -23.9) {$-A^{\star}$};
  \node[txN, anchor=base west, inner sep=0pt] at (77.6, -1.4)  {$\alpha\!=\!0$};
  \node[txN, anchor=base east, inner sep=0pt] at (-69.5, 0.3)  {$\alpha\!=\!\pm\pi$};
  \node[txN, anchor=base west, inner sep=0pt] at (-43.5, 35.6) {$(-\pi,\pi]$};
  \node[txN, anchor=north east, inner sep=1pt] at (-64.5, -29.0) {$\alpha(0)$};
  \node[txP] at (35.0, 30.0) {$\mathcal{P}$};
  \node[txM] at (8:51.5) {$\mathcal{M}$};
  % legend
  \draw[line width=0.9pt, draw={rgb,1:red,0.4392;green,0.4392;blue,0.4392},
        fill={rgb,1:red,0.9372;green,0.9372;blue,0.9372}] (-98.2,-81.3) rectangle (-81.0,-72.7);
  \draw[colP, line width=1.1pt, fill={rgb,1:red,0.8118;green,0.8823;blue,0.9412}]
    (-98.2,-94.0) rectangle (-81.0,-85.5);
  \draw[colM, line width=1.1pt, fill={rgb,1:red,0.9647;green,0.8510;blue,0.7686}]
    (-98.2,-106.8) rectangle (-81.0,-98.3);
  \node[anchor=base west, inner sep=0pt] at (-76.1, -79.6)  {$(-\pi,\pi]$ : all heading errors};
  \node[anchor=base west, inner sep=0pt] at (-76.1, -92.4)  {$\mathcal{P}$ : range decreases monotonically};
  \node[anchor=base west, inner sep=0pt] at (-76.1, -105.2) {$\mathcal{M}\subset\mathcal{P}$};
\end{tikzpicture}
\caption{Schematic for Theorem~\ref{thm:up_attract} showing the evolution of the heading error under the proposed control law with $k>k_{\min}$. The heading error $\alpha(t)$ is driven to the set $\mathcal{P}$ in finite time and approaches the set $\mathcal{M}$ at least exponentially fast thereafter.}
\label{fig:thm_1}
\end{figure}

\medskip

\noindent\emph{(c)} Following parts \emph{(a)} and \emph{(b)}, we conclude that
$|\alpha(t)|<\cos^{-1}\gamma$ for all $t>\tau_{p}$. Lemma \ref{lem:range-hold} further implies
$\dot{\rho}(t)<0$ for $t>\tau_{p}$, regardless of the evader's strategy. Therefore, the range is strictly monotonically decreasing after $t=\tau_p$. It remains to obtain a bound on the capture time $t_c$.

Recall from the proof of Lemma \ref{lem:range-hold} that the maximization over the evader's inputs yields $\dot\rho(t)\le\mu_m-v\cos\alpha(t)$ at every instant and for every
$\alpha$. Integrating this bound over the
entire interval $[0,t_c]$ with $\rho(t_c)=r_c$, we obtain
\begin{align}
& \rho(t_c)-\rho(0) \le \mu_m t_c-v\!\int_{0}^{t_c}\!\cos\alpha(t)\,dt \nonumber \\
\implies & v\!\int_{0}^{t_c}\!\cos\alpha\,dt-\mu_m t_c \;\le\; \rho(0)-r_c \label{eq:bound_int_cos_alpha}.
\end{align}
%%%%%%%%%%%%%%%%%%%%%%%%%%%%%%%%%%%%%%%%%%%%%%%
The trigonometric identity $1-\cos\alpha = 2\sin^2(\frac{\alpha}{2})$, in addition to the inequality $|\sin \alpha|\leq |\alpha|$ for any $\alpha$, yields
\begin{align}
& 1-\cos\alpha
= 2\sin^2\left(\frac{\alpha}{2}\right)
\leq \frac{\alpha^2}{2}
\implies \cos\alpha \geq 1-\frac{\alpha^2}{2} \implies \int_{0}^{t_c}\!\!\cos\alpha\, dt
\geq t_c-\frac{1}{2}\int_{0}^{t_c}\!\!\alpha^{2}\,dt,
\label{eq:bound_vcos_alpha}
\end{align}
where the last inequality follows by integrating the pointwise bound over $[0,t_c]$ and using
$\int_{0}^{t_c}dt=t_c$.

Following \eqref{eq:bound_vcos_alpha}, we obtain
\begin{equation} \label{eq:tau_1}
(v-\mu_m)\,t_c-\frac{v}{2}\!\int_{0}^{t_c}\!\!\alpha^{2}dt \le
v\!\int_{0}^{t_c}\!\!\cos\alpha\,dt-\mu_m t_c .
\end{equation}
Combining \eqref{eq:bound_int_cos_alpha} and \eqref{eq:tau_1}, we get
\begin{equation} \label{eq:tau_2}
(v-\mu_m)\,t_c-\frac{v}{2}\!\int_{0}^{t_c}\!\!\alpha^{2}dt \le \rho(0)-r_c .
\end{equation}
Now, we focus on obtaining a bound on the term $\int_{0}^{t_c}\alpha^{2}dt$.

Since $\frac{d}{dt}\alpha^{2}=2|\alpha|\frac{d}{dt}|\alpha|$, the
inequality \eqref{eq:abs_alpha_dot}, i.e., $\frac{d}{dt}|\alpha|\le-k(|\alpha|-A^{\star})$ implies
$\frac{d}{dt}\alpha^{2}\le-2k\bigl(\alpha^{2}-A^{\star}|\alpha|\bigr)$. Integration of both sides over $[0,t_c]$ yields
\begin{align}
\alpha^{2}(t_c)-\alpha^{2}(0)
&\le -2k\!\int_{0}^{t_c}\!\!\alpha^{2}dt+2kA^{\star}\!\!\int_{0}^{t_c}\!\!|\alpha|\,dt, \nonumber\\
\int_{0}^{t_c}\!\!\alpha^{2}dt
&\le A^{\star}\!\!\int_{0}^{t_c}\!\!|\alpha|\,dt+\frac{\alpha^{2}(0)}{2k},
\label{eq:int_alpha_sq_raw}
\end{align}
where the second inequality follows by rearranging the first, using $-\alpha^{2}(t_c)\le0$
to drop that particular term, and dividing both sides by $2k$. 

Next, the term
$\int_{0}^{t_c}|\alpha|\,dt$ is bounded by integrating
\eqref{eq:abs_alpha_dot} over $[0,t_c]$ and using $|\alpha(t_c)|\ge0$, resulting in
\begin{align}
|\alpha(t_c)|-|\alpha(0)| &\le -k\!\int_{0}^{t_c}\!\!|\alpha|\,dt+kA^{\star}t_c, \nonumber\\
\int_{0}^{t_c}\!\!|\alpha|\,dt &\le A^{\star}t_c+\frac{|\alpha(0)|}{k}. \label{eq:int_abs_alpha}
\end{align}
By substituting the R.H.S. of \eqref{eq:int_abs_alpha} in \eqref{eq:int_alpha_sq_raw}, we obtain
\begin{equation} \label{eq:int_alpha_sq}
\int_{0}^{t_c}\!\!\alpha^{2}dt \;\le\; ({A^{\star}})^{2}t_c
+\frac{A^{\star}|\alpha(0)|}{k}+\frac{\alpha^{2}(0)}{2k}.
\end{equation}
%%%%%%%%%%%%%%%%%%%%%%%
Note that $\int_{0}^{t_c}\alpha^{2}dt$ appears in \eqref{eq:tau_2} with the negative coefficient $-v/2$. We multiply \eqref{eq:int_alpha_sq} with $-v/2$ and add $(v-\mu_m)t_c$ on both sides to obtain
\begin{align}
\delta(k)\,t_c-C
&\le (v-\mu_m)\,t_c-\frac{v}{2}\!\int_{0}^{t_c}\!\!\alpha^{2}dt
\le \rho(0)-r_c, \label{eq:tau_s_chain}
\end{align}
where the second inequality follows from \eqref{eq:tau_2}, and $\delta(k)$ and $C$ are as defined in \eqref{eq:delta} and \eqref{eq:Cconst}, with $C\ge0$ since $A^{\star}>0$. 

The coefficient $\delta(k)$ is positive if and only if $k^{2}>v\mu_m^{2}/\bigl(2r_c^{2}(v-\mu_m)\bigr)$, that is, if and only if $k>k_{\min}$, where $k_{\min}$ is the prescribed minimum gain from \eqref{eq:kmin_def}. Thus, upon rearranging the first and last inequalities in \eqref{eq:tau_s_chain} and dividing both sides by $\delta(k)>0$, we obtain the bound
\begin{equation*}
  t_c \;\le\; \bar{t}_c \;=\; \frac{\rho(0)-r_c+C}{\delta(k)},
\end{equation*}
as prescribed in \eqref{eq:tcbound}. Note that $k_{\min}$ is finite when $r_c>0$. Thus, the evader is always captured no later than $t=\bar{t}_c$ regardless of its strategy. 

Finally, we show that $\bar{\tau}_p < \bar{t}_c$.
%%%%%%%%%%%%%%%%%%%%%%%%%%%%%%%%%%%%%%%%%%%%%%%%%%%%%%
If $|\alpha(0)|\le\cos^{-1}\gamma$, then $\bar\tau_p=0$ by \eqref{eq:tauP} and the claim is trivial. We therefore consider the case $|\alpha(0)|>\cos^{-1}\gamma$. Since $v-\mu_m=v(1-\gamma)=\frac{v}{2}s^{2}$, where $s:=\sqrt{2(1-\gamma)}$, and
$v\mu_m^{2}/(2k^{2}r_c^{2})=\frac{v}{2}(A^{\star})^{2}$, the coefficient $\delta(k)$ in
\eqref{eq:delta} factorizes as
\begin{equation}\label{eq:delta_fact}
  \delta(k)=\frac{v}{2}\bigl(s^{2}-(A^{\star})^{2}\bigr)
  =\frac{v}{2}(s-A^{\star})(s+A^{\star}).
\end{equation}
Furthermore, the relationship $A^{\star}<s<\cos^{-1}\gamma$ established in part \emph{(b)} yields
$s-A^{\star}\le\cos^{-1}\gamma-A^{\star}$ and $s+A^{\star}<2\cos^{-1}\gamma$. Consequently, from \eqref{eq:delta_fact}, we obtain
\begin{equation*}
  4\delta(k) \;\le\; 4v\,\cos^{-1}\gamma\,\bigl(\cos^{-1}\gamma-A^{\star}\bigr).
\end{equation*}

We multiply both sides by $|\alpha(0)|-\cos^{-1}\gamma>0$ to obtain
\begin{equation*}
  4\delta(k)\bigl(|\alpha(0)|-\cos^{-1}\gamma\bigr)
  \;\le\; v\,\bigl(\cos^{-1}\gamma-A^{\star}\bigr)\,\alpha^{2}(0),
\end{equation*}
where the inequality holds because $\bigl(|\alpha(0)|-2\cos^{-1}\gamma\bigr)^{2}\ge0$, which can be rearranged to obtain $4\cos^{-1}\gamma\bigl(|\alpha(0)|-\cos^{-1}\gamma\bigr)\le\alpha^{2}(0)$. Dividing both sides by $4k\,\delta(k)\bigl(\cos^{-1}\gamma-A^{\star}\bigr)>0$ yields
\begin{align*}
    & \bar\tau_p = \frac{|\alpha(0)|-\cos^{-1}\gamma}{k (\cos^{-1}\gamma-A^{\star})}
  \;\le\; \frac{v\,\alpha^{2}(0)}{4k\,\delta(k)} \;\le\; \frac{C}{\delta(k)} \;<\; \bar t_c,
\end{align*}
which follows from \eqref{eq:Cconst} and $\rho(0)>r_c$. This completes the proof.
\end{proof}

%%%%%%%%%%%%%%%%%%%%%%%%%
%%%%%%%%%%%%%%%%%%%%%%
%%%%%%%%%%%%%%%%%%%%%%
\begin{remark}
Note that Theorem \ref{thm:up_attract} is applicable when the set of admissible gains $(k_{\min},k_{\max}]$ is non-empty, or equivalently, 
\begin{equation}\label{eq:omega_feas}
  k_{\max} > k_{\min} \iff \omega_m \;>\; k_{\min}\pi+\frac{v}{r_c},
\end{equation}
where $k_{\min}$ depends only on $v$, $\mu_m$, and $r_c$ through
\eqref{eq:kmin_def}. In other words, whenever the pursuer's parameters $v$, $\omega_m$, and
$r_c$, together with the known evader speed bound $\mu_m<v$, satisfy
\eqref{eq:omega_feas}, the admissible gain interval is non-empty, and every
gain $k\in(k_{\min},k_{\max}]$ guarantees finite-time capture while
maintaining $|\omega_p|\le\omega_m$ throughout the engagement.
\end{remark}
%%%%%%%%%%%%%%%%%%%%%%%%%%%%%%%%%%%%%%%%%%%%%%%%%

\begin{remark}%[Relation to the classical homicidal chauffeur game]
\label{rem:hcg_regime}
Let $R_{\min}:=v/\omega_m$ be the minimum turning radius and $\ell:=r_c/R_{\min}$ be the ratio of the capture radius to the minimum turning radius, i.e., the normalized capture radius \cite{ref_isaacs1999differential}. Since $k_{\min}>0$, the non-emptiness condition \eqref{eq:omega_feas} gives $\omega_m>v/r_c$, thus resulting in $\ell>1$. As a result, we get $\gamma^2+\ell^2>1$ with $\gamma=\mu_m/v$. In contrast, most works on classical HCG \cite[Sec. 3.2]{ref_hcg_pachter2019classical}, \cite{ref_mahapatra2025deception} are posed in the complementary regime $\gamma^2+\ell^2<1$.
\end{remark}
%}
%%%%%%%%%%%%%%%%%%%%%%%
Condition \eqref{eq:omega_feas} is a joint requirement on the pursuer's agility, its capture radius, and its speed advantage. We now consider the case when the bound on the evader's speed $\mu_m$ is unknown to the pursuer. The following proposition identifies a speed ceiling $\mu^\star$, computable from the pursuer's own fixed parameters $v$, $\omega_m$, and $r_c$ alone, below which the gain interval remains non-empty and the choice $k=k_{\max}$ preserves the guarantees of
Theorem \ref{thm:up_attract}.

%%%%%%%%%%%%%%%%%%%%%%%%%%%%%%%%%%%%%%%%%%%%%%%
\begin{proposition}[Certified speed ceiling for unknown $\mu_m$]\label{prop:mustar}
Suppose $\omega_m > v/r_c$, so that $k_{\max}>0$ in \eqref{eq:k_max}, and let $k_{\min}(\mu)$, $\delta(k;\mu)$ and $C(k;\mu)$ be as defined in \eqref{eq:kmin_def}, \eqref{eq:delta}, and \eqref{eq:Cconst}, respectively, with $\mu_m$ replaced by $\mu\in(0,v)$, which is treated as a variable. Let $\mu^\star$ denote the unique root of the equation 
$k_{\min}(\mu)=k_{\max}$ in $(0,v)$. Then, the gain interval $(k_{\min}(\mu_m),k_{\max}]$ is non-empty if and only if $\mu_m<\mu^\star$. Furthermore, for every $\bar\mu<\mu^\star$, the bound
\begin{equation}\label{eq:tc_uniform}
  t_c \;\le\; \frac{\rho(0)-r_c+C(k_{\max};\bar\mu)}{\delta(k_{\max};\bar\mu)}
\end{equation}
holds uniformly for every evader with $\mu_m\le\bar\mu$.
\end{proposition}
%%%%%%%%%%%%%%%%%%%%%%%%%%%%%%%%%%%%%%%%%%%%%%%%%%%%%%%%%%%%%%%%%%%%%%%%%%%%
\begin{proof}
Since $k_{\max}>0$ and $k_{\min}(\mu)\ge0$, the equation
$k_{\min}(\mu)=k_{\max}$ is equivalent to $k_{\min}(\mu)^2=k_{\max}^2$, which yields
\begin{equation}\label{eq:mustar}
  v\,\mu^{2} + 2k_{\max}^{2}r_c^{2}\,\mu - 2k_{\max}^{2}r_c^{2}\,v = 0 .
\end{equation}
The L.H.S. is a quadratic function of $\mu$ whose value is strictly positive when $\mu=v$ and is strictly negative when $\mu=0$. Furthermore, it is
strictly increasing for $\mu \in (0,v)$, since its derivative, $2v\mu+2k_{\max}^{2}r_c^{2}$, is positive in this range. Therefore, there exists a unique root in the interval $(0,v)$ which is denoted by $\mu^\star$. At $\mu=\mu^\star$ we have $k_{\min}(\mu^\star)=k_{\max}$ by definition, and therefore, the gain interval is empty, and Theorem \ref{thm:up_attract} is not applicable. 

Let us now fix $\bar\mu<\mu^\star$ and set $k=k_{\max}$. Then, for any $\mu_m \leq \bar\mu$, we have $k_{\min}(\mu_m)\le k_{\min}(\bar\mu)<k_{\max}$ since $k_{\min}(\mu)$ is a monotonically increasing function of $\mu$. Thus, Theorem \ref{thm:up_attract} is applicable and $\delta(k_{\max};\mu_m)>0$. 
Observe that $\delta(k_{\max};\mu)$ is strictly decreasing in $\mu$ while $C(k_{\max};\mu)$ is non-decreasing in $\mu$ because $A^{\star}=\mu/(k_{\max}r_c)$. Therefore, the bound \eqref{eq:tcbound} is non-decreasing in $\mu$, and evaluating it at $\mu=\bar\mu$ yields \eqref{eq:tc_uniform}. 
\end{proof} 

To summarize, when the pursuer selects $k=k_{\max}$ without any knowledge of the evader's actual speed bound $\mu_m$, Theorem \ref{thm:up_attract} guarantees finite-time capture for every evader with $\mu_m<\mu^\star$ while respecting the turn constraint.

%%%%%%%%%%%%%%%%%%%%%%%%%%%%%%%%%%%%%%%%%%%%%%%%%%%%%%%%%%%% 
%%%%%%%%%%%%%%%%%%%%%%%%%%%%%%%%%%%%%%%%%%%%%%%%%%%%%%%%%%%% 
\section{Numerical Results}\label{section:numerical}
%\textcolor{blue}{Rewriting in progress}
%%%%%%%%%%%%%%%%%%%%%%%%%%%%%%%%%%%
We now demonstrate the effectiveness of the proposed strategy via numerical simulations carried out in the MATLAB environment. The capture radius is taken as $r_c=0.125 \,\text{m}$. We consider a pursuer with speed $v=0.5$\,m/s and turn-rate limit $\omega_m=10$\,rad/s, so that its minimum turning radius is $R_{\min}=v/\omega_m=0.05$\,m.
The evader is considered to move at its maximum speed, which is set to be $\mu_m=0.2$\,m/s.  Considering these values, the controller gains from \eqref{eq:kmin_def} and \eqref{eq:k_max} are computed to be $k_{\min}=1.46$ and $k_{\max}=1.91$, respectively, so the condition \eqref{eq:omega_feas} holds and we set $k=k_{\max}$. For the chosen parameters, Proposition \ref{prop:mustar} gives $\mu^\star=0.242$\,m/s from \eqref{eq:mustar}. Since $k=k_{\max}$ is chosen from the pursuer's parameters alone, the capture guarantee of Theorem \ref{thm:up_attract} holds for every evader with $\mu_m<\mu^\star$, without the pursuer using the value of $\mu_m$.

The agents are initialized at $p(0) = (0,\,0) \,\text{m}$ and $e(0) = (-0.186,\,-0.026) \,\text{m}$, giving an initial range $\rho(0) = 0.188 \,\text{m}$. The initial heading angle of the pursuer is chosen to be $\theta(0)=0$, which results in  $\beta(0)=\alpha(0)=-3.00 \approx -\pi$\,rad, i.e., the pursuer initially faces almost directly away from the evader. For this configuration, $\delta(k_{\max})=0.125$ and $C=0.919$ from \eqref{eq:delta} and \eqref{eq:Cconst}, and thus \eqref{eq:tauP} and \eqref{eq:tcbound} give the certified bounds $\bar\tau_p=3.00$\,s and $\bar t_c=7.89$\,s, respectively.

\subsection{Representative Evader Policies}
We consider the following evader policies, which span qualitatively distinct ways of delaying capture, while the pursuer applies \eqref{eq:thm_controller} with $k=k_{\max}$.
\begin{enumerate}
 \item \textbf{LOS-normal evasion}: The evader moves orthogonally to the LOS, i.e., $\psi(t)=\beta(t)+\tfrac{\pi}{2}\,\mathrm{sgn}\big(\alpha(t)\big)$. This
maximizes the evader term $d(t)$ in \eqref{eq:abs_alpha_dot} and is therefore the policy that directly opposes the convergence established in Theorem \ref{thm:up_attract}. Although $\psi(t)$ is discontinuous at $\alpha=0$, this policy is admissible under Assumption \ref{ass:evader_strategy}, and the guarantees of Theorem \ref{thm:up_attract} therefore continue to hold.
\item \textbf{Random evasion}: 
 The evader's heading is commanded online by a human operator to emulate unpredictable and erratic behavior \cite{ref_RMPC_patra20253d,ref_encirclement_patra2026encirclement,ref_slow_encirclement_wang2023distributed}. This policy admits no analytical model and is included to demonstrate that the proposed guarantee remains valid against an adversary whose behavior is not known to the designer.
 \item \textbf{HCG evasion}: The evader plays the continuous-time feedback equilibrium strategy of the homicidal chauffeur game (HCG) \cite{ref_isaacs1999differential,ref_hcg_merz1974homicidal,ref_hcg_pachter2019classical} against our proposed strategy. Here, the normalized capture radius \cite{ref_isaacs1999differential} is $\ell=r_c/R_{\min}=r_c\omega_m/v=2.5$. For such parameters the HCG admits no barrier, and therefore no equivocal surface or secondary trajectories \cite{ref_hcg_pachter2019classical}, \cite[p. 439]{BasarOlsder1999}. Furthermore, for $\ell\ge2$, the equilibrium reduces to a hard turn of the pursuer followed by a straight chase \cite[p. 75]{ref_hcg_merz1974homicidal}. The evader then runs along the tangent from its position to the pursuer's minimum-turn circle on the side toward which the pursuer turns \cite{ref_isaacs1999differential}, and near the capture circle its heading follows the primary trajectories \cite[Sec. 3.1]{ref_hcg_pachter2019classical}.
\end{enumerate}

As a benchmark, we also simulate the \emph{HCG pursuit game}, in which the pursuer plays its HCG equilibrium strategy, that is, it turns at $\pm\omega_m$ toward the evader until $\alpha=0$ and then moves straight along the LOS. This benchmark gives the min-max capture time for the same initial configuration.

%%%%%%%%%%%%%%%%%%%%%%%%%%%%%%%%%%%%%%%%%%%%%%%%%%%%%%%
\begin{figure*}[t!]
      \centering
      \begin{subfigure}{0.24\textwidth}
       \centering 
     \includegraphics[width=\textwidth]{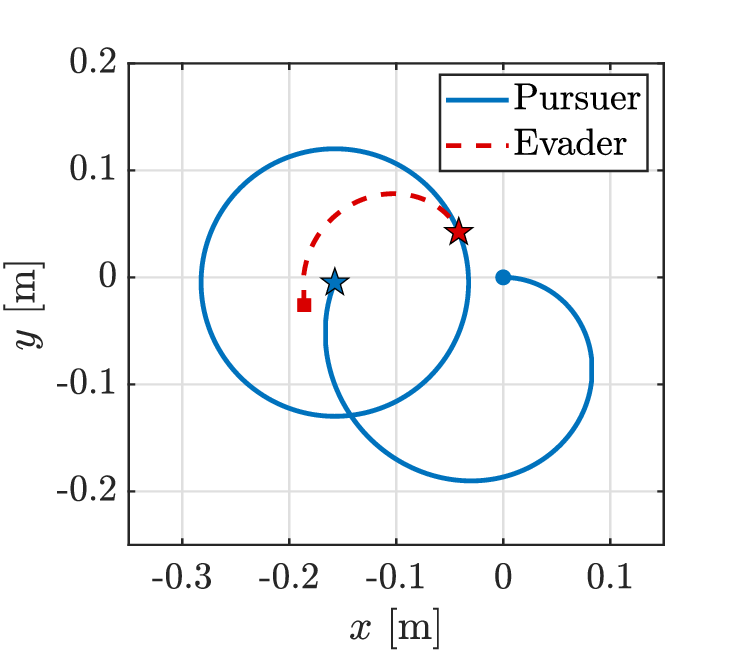}
     \caption{LOS-normal evasion}
      \label{fig2:perp}
      \end{subfigure}
    \begin{subfigure}{0.24\textwidth}
       \centering 
     \includegraphics[width=\textwidth]{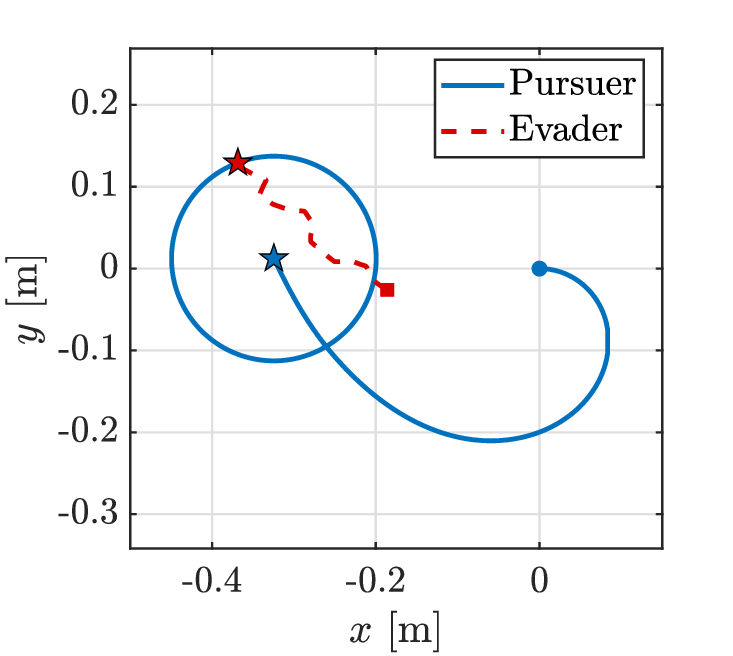}
     \caption{Random evasion}
      \label{fig3:random}
      \end{subfigure}
     \begin{subfigure}{0.24\textwidth}
       \centering 
     \includegraphics[width=\textwidth]{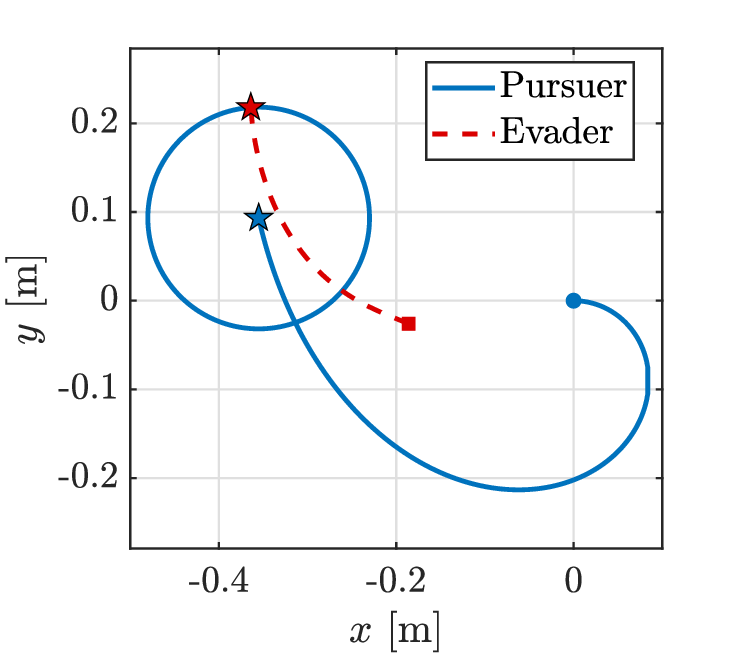}
     \caption{HCG evasion}
      \label{fig3:hcg_evasion}
      \end{subfigure}
      %%%
    \begin{subfigure}{0.24\textwidth}
       \centering 
     \includegraphics[width=\textwidth]{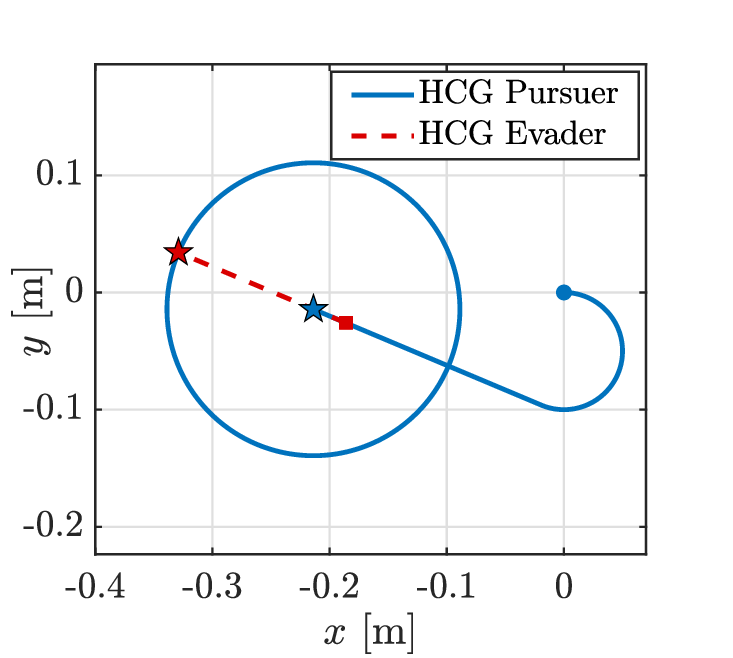}
     \caption{HCG game}
      \label{fig4:hcg_game}
      \end{subfigure}
       \centering
      \caption{Evolution of the pursuer and the evader trajectories under the
representative evader policies and the HCG game in which both players use
their equilibrium strategies. The pursuer's and evader's initial positions are
marked by a filled circle and a filled square, respectively, and their terminal positions at capture by stars. In each case, the game terminates with the evader getting captured at
$t_c=1.14$\,s (LOS-normal evasion), $1.42$\,s (Random evasion), and $1.61$\,s (HCG evasion), against the certified bound $\bar t_c=7.89$\,s. In the HCG game, the evader gets captured comparatively faster at $0.78$\,s.}
      \label{fig:pp_unicycle}
   \end{figure*}
%%%%%%%%%%%%%%%%%%%%%%
\begin{figure*}[t!]
      \centering
      \begin{subfigure}{0.24\textwidth}
       \centering 
     \includegraphics[width=\textwidth]{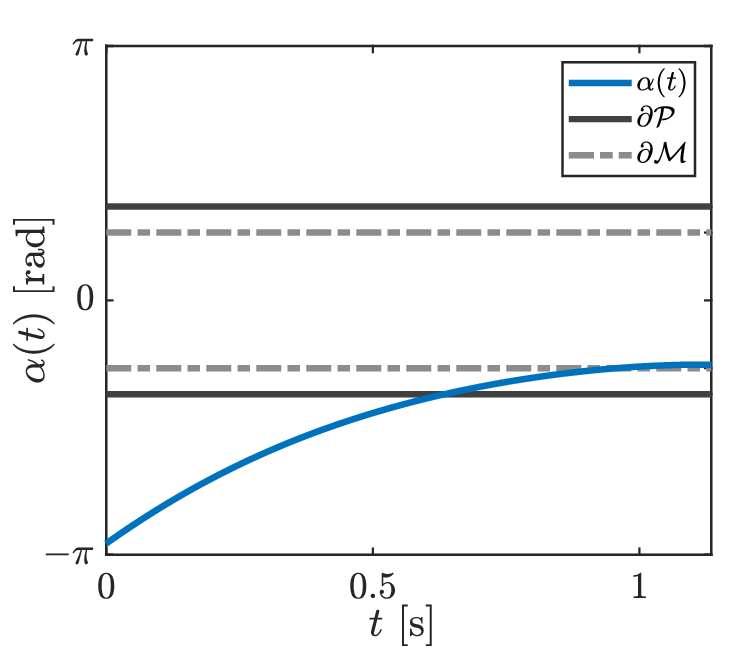}
     \caption{LOS-normal evasion}
      \label{fig1:alpha_perp}
      \end{subfigure}
    \begin{subfigure}{0.24\textwidth}
       \centering 
     \includegraphics[width=\textwidth]{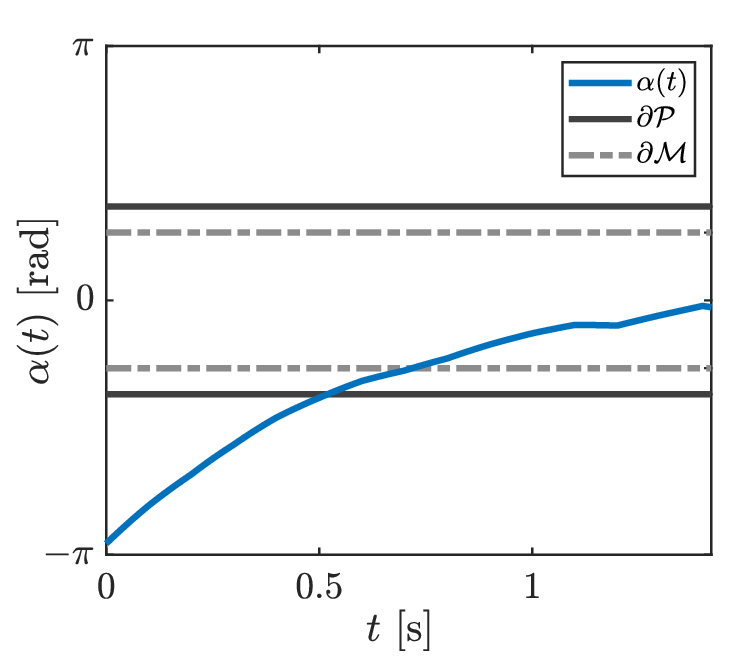}
     \caption{Random evasion}
      \label{fig2:alpha_random}
      \end{subfigure}
    \begin{subfigure}{0.24\textwidth}
       \centering 
     \includegraphics[width=\textwidth]{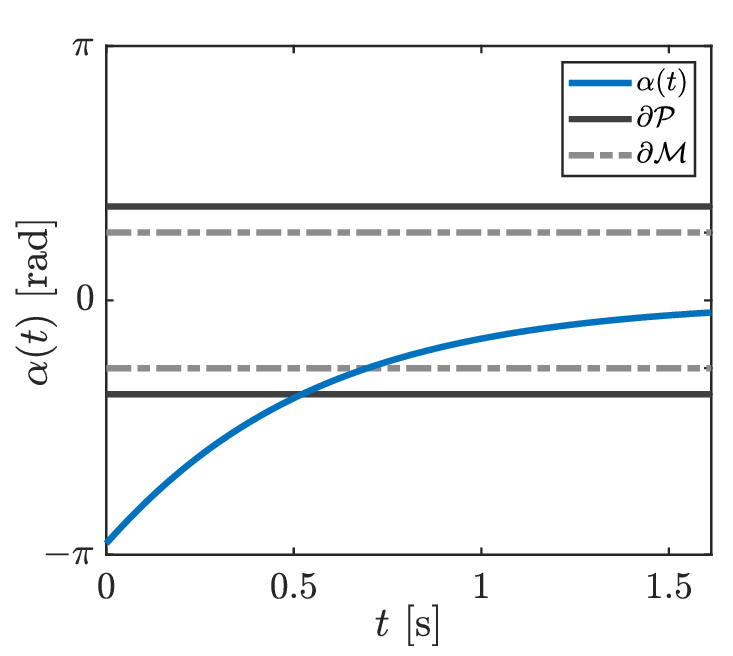}
     \caption{HCG evasion}
      \label{fig3:alpha_hcg}
      \end{subfigure}
      %%%
      \begin{subfigure}{0.24\textwidth}
       \centering 
     \includegraphics[width=\textwidth]{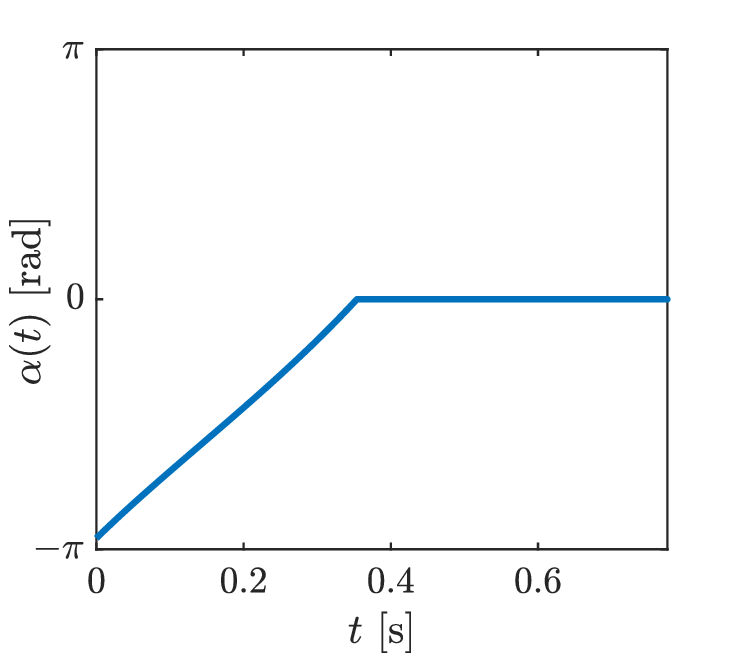}
     \caption{HCG game}
      \label{fig4:alpha_hcg_game}
      \end{subfigure}
      %%%
       \centering
      \caption{Evolution of the heading error $\alpha(t)$ under the representative evader policies and the HCG pursuit game. The boundary $\partial\mathcal{P}$ of the set $\mathcal{P}$ is shown by solid black lines and the boundary $\partial\mathcal{M}$ of the set
       $\mathcal{M}$ by grey dash-dotted lines. The entry times into $\mathcal{P}$ are
       $\tau_p = 0.634$\,s (LOS-normal Evasion), $0.519$\,s (Random evasion), and $0.522$\,s (HCG evasion), against the certified bound $\bar{\tau}_p = 3$\,s. In the HCG game, $\alpha$ decreases in magnitude almost linearly and reaches $\alpha=0$ at around $0.35$\,s.}
     \label{fig:alpha_traj}
     \end{figure*}
\begin{figure}[t!]
    \centering
    \includegraphics[width=0.48\linewidth]{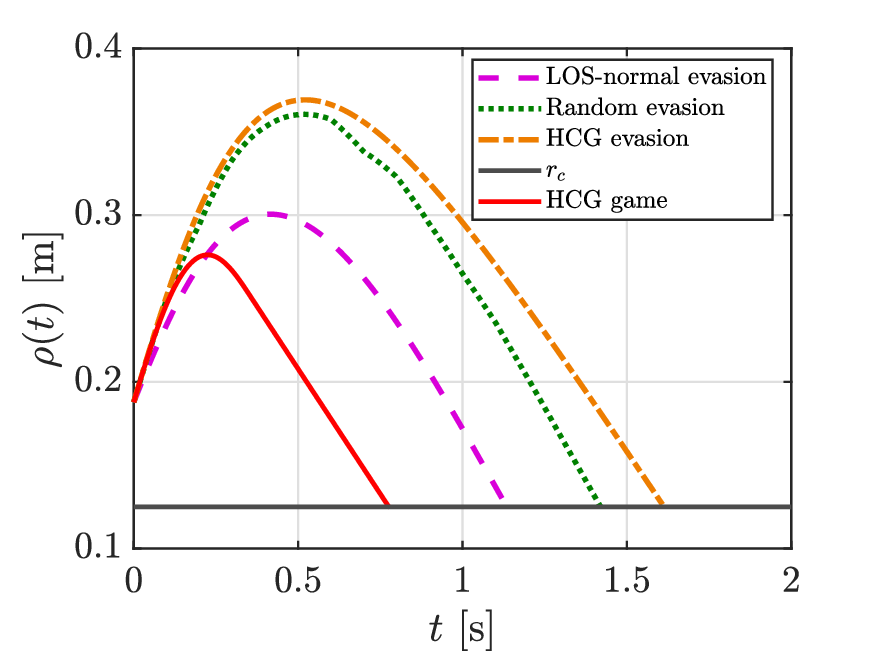}
    \caption{$\rho(t)$ under the representative evader policies and in the HCG
pursuit game, from a common initial configuration. The horizontal line marks
the capture radius $r_c=0.125$\,m. In all cases, the evader is captured well
within the bound $\bar t_c=7.89$\,s certified by
Theorem \ref{thm:up_attract}. Against the same HCG evader, the time-optimal HCG pursuer captures faster ($0.78$\,s) than the proposed law ($1.61$\,s).}
    \label{fig:dist_comp}
\end{figure}
\begin{figure}[ht!]
    \centering
    \includegraphics[width=0.48\linewidth]{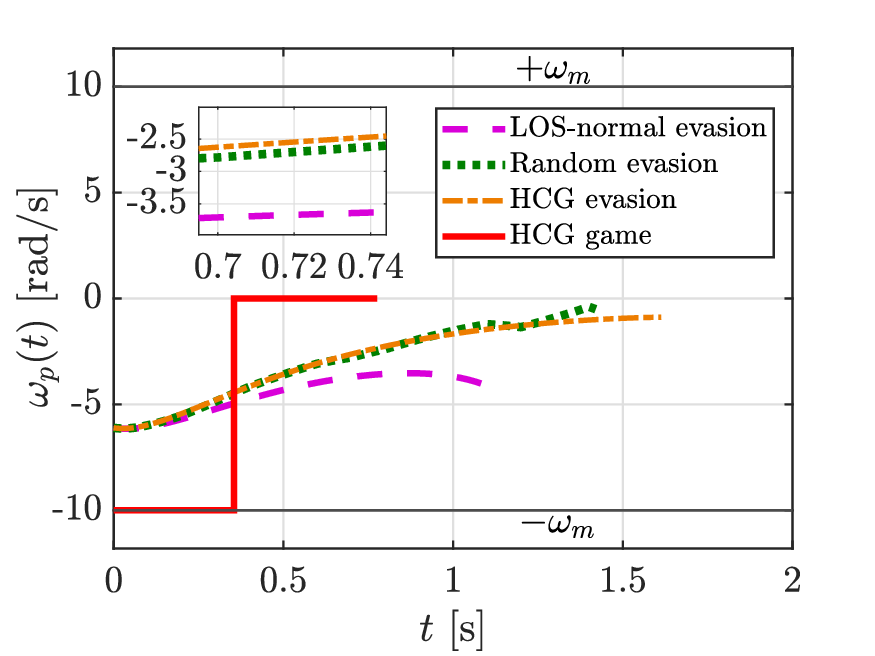}
    \caption{Pursuer turn rate $\omega_p(t)$ under the representative evader
policies and in the HCG pursuit game, with the actuator limits
$\pm\omega_m=\pm10$\,rad/s shown as horizontal lines. Under the proposed law,
$|\omega_p|\le\omega_m$ holds at all times, with a peak of $6.1$\,rad/s due to
the anti-aligned initial heading. The HCG pursuer's command is discontinuous, i.e., it saturates at $-\omega_m$ until $\alpha=0$ and then switches to $\omega_p=0$.}
    \label{fig:w_comp}
\end{figure}

%%%%%%%%%%%%%%%%%%%%%%%%%%%%%%%%%%%%%%%%%%%%%%%%%%%%%%%
The kinematics of the pursuer \eqref{eq:P_dyn} in closed loop under the proposed control law \eqref{eq:thm_controller} and of the evader \eqref{eq:E_dyn} for each case are  integrated in MATLAB using the \texttt{ode45} solver. Figures \ref{fig:pp_unicycle}--\ref{fig:w_comp} show the trajectories, the heading error, the range, and the pursuer's turn rate. Figure \ref{fig:pp_unicycle} shows that in every case the engagement terminates with the evader inside the pursuer's capture disc. Under the LOS normal evasion policy shown in Fig. \ref{fig2:perp}, the evader breaks orthogonally to the LOS, forcing the pursuer into a tight curved capture arc. Capture still follows under this policy after a single turn. Under random evasion shown in Fig. \ref{fig3:random}, the evader's trajectory is erratic since its heading is controlled online manually, yet the pursuer tracks it throughout, and the capture is finally attained. Under HCG evasion (Fig. \ref{fig3:hcg_evasion}), the evader exploits the pursuer's turning phase by running along the tangent from its position to the pursuer's
minimum-turn circle; nevertheless, it is finally captured. Finally, in the HCG game (Fig. \ref{fig4:hcg_game}), the pursuer turns at full rate and then chases the evader straight along the LOS to finally capture it.

Figure \ref{fig:alpha_traj} shows the evolution of the heading error under the proposed law for LOS normal, random, and HCG evasion and in the HCG pursuit game. In every case
$\alpha$ starts near $-\pi$, since the evader is initially behind the pursuer.
Under the proposed law, $\alpha$ is driven into $\mathcal P$ at $\tau_p = 0.634$\,s (LOS-normal evasion), $0.519$\,s (Random evasion), and $0.522$\,s (HCG evasion), against the certified bound $\bar{\tau}_p = 3$\,s. Thereafter $\alpha$ remains in
$\mathcal P$ and converges toward $\mathcal M$ at least as fast as an exponential rate, as guaranteed in Theorem \ref{thm:up_attract}(a)--(b). In the HCG pursuit game, the pursuer turns at full rate, so $\alpha$ decreases in magnitude almost linearly, and reaches $\alpha=0$ at $0.35$\,s, where it is held until capture.  
%%%%%%%%%%%%%%%%%%%%%%%%%%%%%%%%%%%%%%%%%%%%%%%%%%%%%%%%%%%%%%%%%%%%%

Figure \ref{fig:dist_comp} shows the range. Since the evader starts behind the
pursuer, $\alpha(0)\notin\mathcal P$, and by Lemma \ref{lem:range-hold} the evader can initially increase the range. Consequently, $\rho$ first grows while the pursuer turns,
and decreases once $\alpha$ enters $\mathcal P$. 
Under the proposed law, HCG evasion yields the longest capture time ($1.61$\,s), followed by random evasion ($1.42$\,s), while LOS-normal evasion yields the shortest ($1.14$\,s).
All of these are well within the certified bound $\bar t_c=7.89$\,s. As expected, the HCG pursuit game yields a faster capture ($0.78$\,s), as the HCG pursuer is time-optimal. The HCG pursuer gains this advantage by turning at its maximum rate, which
requires a saturated, discontinuous turn-rate command, as discussed next.
%%%%%%%%%%%%%%%%%%%%%%%%%%%%%%%%%%%%%%%%%%%%%%%%%%%%%%%%%%%%%%%
%%%%%%%%%%%%%%%%%%%%%%%%%%%%%%%%%%%%%%%%%%%%%%%%%%%%%%%%%%%%%%%

Figure \ref{fig:w_comp} shows that under the proposed law,
the constraint $|\omega_p|\le\omega_m$ holds at all times for every evader policy. 
The high initial effort is due to the anti-aligned initial heading,
yet the command peaks at only $6.1$\,rad/s, i.e., $61\%$ of $\omega_m$, and varies continuously in time without switching since $\omega_p$ is continuous by Theorem \ref{thm:up_attract}(a). In contrast, the HCG pursuer's turn rate saturates at $-\omega_m$ for the entire turn and then switches abruptly to $\omega_p=0$ at $t=0.35$\,s. The proposed law thus trades a longer capture time for a smooth, unsaturated turn-rate input, together with closed-form certificates on the entry and capture times that hold against every admissible evader.

\subsection{Comparison of Numerical Capture Time with the Certified Bound}
Next, we validate the capture-time bound \eqref{eq:tcbound} by performing Monte
Carlo simulations with $N = 100$ trials under the HCG evasion policy of the
evader, which is observed to be the worst case for the capture time among the policies considered. In each trial, the pursuer is initialized at its fixed position $p(0) = (2,0)$ m with heading $\theta(0) = 0$, while the evader's initial position is drawn uniformly at random from an annulus of radii $0.5$ m and $1$ m centered at the pursuer, thereby randomizing both arguments of the bound, namely the initial range $\rho(0)$ and the initial heading error $\alpha(0)$. For each trial, we compute (i) the actual capture time $t_{\mathrm{sim}}$ by numerically simulating the closed-loop dynamics under the proposed control law \eqref{eq:thm_controller}, and (ii) the theoretical upper bound $\bar{t}_c$ from \eqref{eq:tcbound}. 

\begin{figure}[htpb!]
      \centering
      \begin{subfigure}[t]{0.48\textwidth}
       \centering 
         \includegraphics[width=\linewidth]{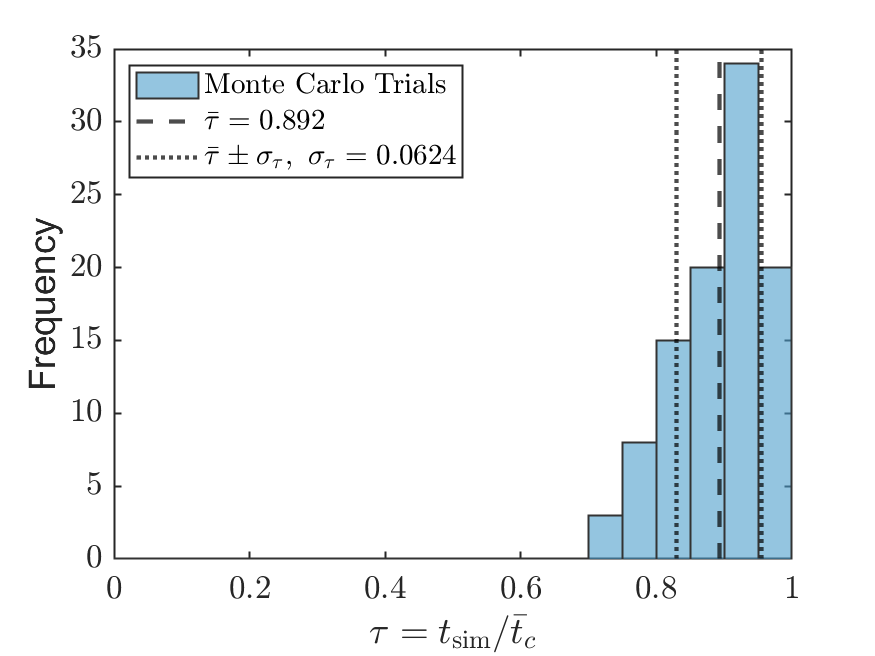}   
         \caption{$\mu_m=0.05$ m/s}
         \label{fig:mu_pt05}
      \end{subfigure} \hfill
      %% C
      \begin{subfigure}[t]{0.48\textwidth}
       \centering 
     \includegraphics[width=\linewidth]{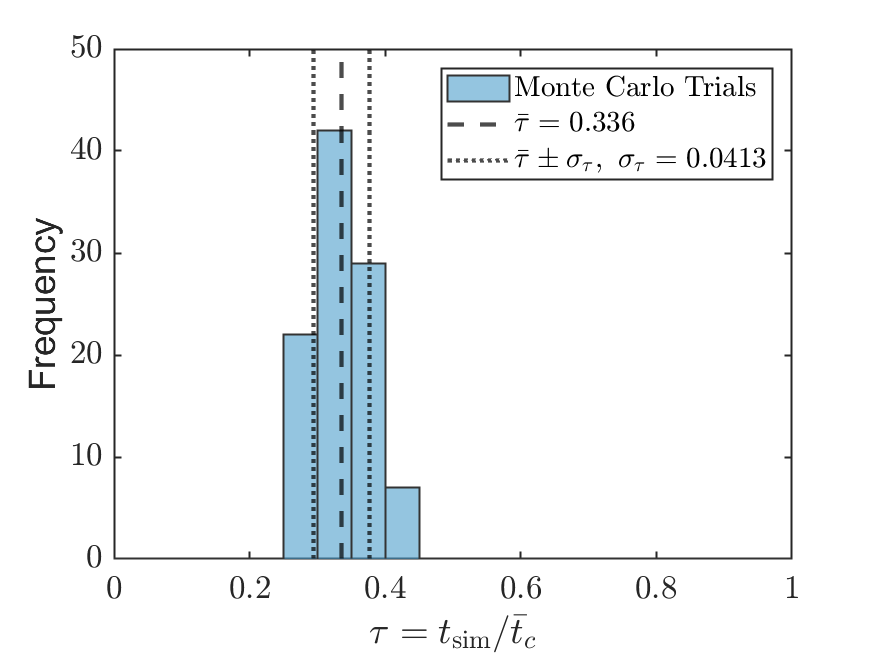}
     \caption{$\mu_m=0.2$ m/s}
      \label{fig:mu_pt2}
      \end{subfigure}
      %%
      %%
      %%%
      \caption{Empirical distribution, mean, and standard deviation of the ratio $\tau$ of the numerical capture time $t_{\mathrm{sim}}$ to the theoretical upper bound $\bar{t}_c$ obtained from \eqref{eq:tcbound}, under the HCG evasion policy of the evader with the pursuer initialized at $p(0) = (2,0)$~m with $\theta(0) = 0$ and $v=0.5$ m/s, while the evader's initial position is drawn uniformly at random from an annulus centered at the pursuer. It is found that $\tau < 1$ ($t_{\mathrm{sim}} < \bar{t}_c$) in each case, and that the bound is more conservative for $\mu_m = 0.2$ m/s compared to $\mu_m = 0.05$ m/s in line with the theoretical findings.}
      \label{fig:capture_mc}
\end{figure}

The empirical distributions of the ratio $\tau = t_{\mathrm{sim}}/\bar{t}_c$ are shown
in Fig. \ref{fig:mu_pt05} for $\mu_m = 0.05$ m/s and Fig. \ref{fig:mu_pt2} for
$\mu_m = 0.2$ m/s, considering $v = 0.5$ m/s. 
We keep the values of $v$, $r_c$, and $\omega_m$ unchanged, and thus
%We take the same values $v = 0.5$ m/s, $r_c = 0.05$ m, and $\omega_m = 130$ rad/s,
$k_{\max} = 1.91$ remains unchanged, and is independent of
$\mu_m$. The binding lower limit is the one for $\mu_m = 0.2$ m/s, namely $k_{\min} = 1.46$, against $k_{\min} = 0.30$ for $\mu_m = 0.05$ m/s.
We therefore set $k = k_{\max}$, which is admissible in both cases and taken as common so that the comparison isolates the effect of $\mu_m$.  
Figure \ref{fig:capture_mc} shows that $\tau < 1$ is always satisfied, i.e., $\bar{t}_c$ exceeds $t_{\mathrm{sim}}$ in every trial. Furthermore, the bound is close to tight for $\mu_m = 0.05$ m/s but considerably more conservative for $\mu_m = 0.2$ m/s.
This follows from the denominator $\delta(k)$ in \eqref{eq:delta}, which becomes
small as $\mu_m$ increases. The bound $\bar{t}_c$ therefore grows sharply, while $t_{\mathrm{sim}}$ does not and $\tau$ decreases accordingly.

\section{Conclusion}\label{section:conclusion}
We considered a planar pursuit-evasion problem in which a constant-speed Dubins pursuer with a bounded turn rate seeks to capture a point-mass evader with first-order kinematics, using only the evader's position and an upper bound on its speed. We proposed a turn-rate control law with a single tunable gain and derived the range of gains for which finite-time capture is guaranteed irrespective of the evader's strategy, while the prescribed turn-rate limit is respected. This range is non-empty if and only if the pursuer's turning authority exceeds an explicit threshold determined by its speed, the evader's speed bound, and the capture radius. Furthermore, we obtained a closed-form upper bound on the capture time, valid against every admissible evader strategy. 
Finally, we showed that the approach extends to the case of an unknown evader speed bound, i.e., the guarantees on capture and turn constraint satisfaction hold against every evader whose speed bound lies below a ceiling computable from the pursuer's parameters alone.

Several directions remain open for future work. It is of interest to derive a tighter capture-time bound in the future.  Extending the setting to multiple nonholonomic pursuers in a cluttered environment, with obstacle and inter-pursuer collision avoidance, also remains an important direction for future research.

\section*{Acknowledgments}
The authors thank Shaunak D. Bopardikar and Dipankar Maity for useful discussions. 
Claude Opus 5.5 (Anthropic) was used to assist with language editing and proofreading of the manuscript. All technical content, derivations, and results are the authors' own, and the authors reviewed all AI-assisted material and take full responsibility for the manuscript.

\bibliography{main}

\end{document}